# Space-Time Nonseparable Electromagnetic Vortices

Minhui Xiong, Ren Wang, Qinyu Zhou, Bing-Zhong Wang

Institute of Applied Physics, University of Electronic Science and Technology of China, Chengdu 611731, China

Corresponding author: Ren Wang; e-mail: rwang@uestc.edu.cn

## Abstract

In structured light with controllable degrees of freedom (DoFs), the vortex beams carrying orbital angular momentum (OAM) give access to provide additional degrees of freedom for information transfer, and in classic field, the propagation-invariant space–time electromagnetic pulses are the possible approach to high-dimensional states. This paper arose an idea that coupling the space-polarization nonseparable states of vortex beams and space-time nonseparable states of spatiotemporal pulse can generate numerous unique and beneficial effects. Here, we introduce an family of space-time nonseparable electromagnetic vortices (STNEV). The pulses exhibit complex and robust spatiotemporal topological structure of the electromagnetic fields, multiple singularities in the Poynting vector maps and distributions of energy backflow. We apply a quantum-mechanics methodology for quantitatively characterizing space-time nonseparability of the pulse. Our findings facilitate their applications in fields of information transfer, toroidal electrodynamics and inducing transient excitations in matter.

## Introduction

Structured light with controllable degrees of freedom has fuelled many applications, from communication, metrology, microscopy, imaging, quantum processing to light–matter interactions[1-3]. Vortex beams, a kind of important structured light, have attracted extensive attention due to their orbital angular momentum (OAM)[4,5], singularities[6], and applications[7-9]. Mathematically analogue to quantum entangled states, classical nonseparable states of structured light were recently hatched attractive studies[10]. Based on the qubit Bell states[11,12], the entanglement of spin angular momentum (SAM) and OAM has been realized to beating communication capacity limit by using the space-polarization nonseparable states of vortex beams[13-17].

In quantum optics, high-dimensional Bell states can be realized[18,19]. However, such states cannot be realized based on two-dimensional polarization degree of freedom in classical[10]. Selection two degrees of freedom from spatiotemporal structured light pulses, i.e. space and time (or equivalently frequency or wavelength), is an alternative route to classical high-dimensional Bell states. Recently, propagation-invariant space–time wave packets with anomalous refraction and arbitrary group velocities were studied[20-22]. The propagation-invariant distance of a space–time wave packets is related to the degree of space-time nonseparability[23,24]. In 1989, Ziolkowski discovered a class of space-time nonseparable pulse solutions of Maxwell's equations, named "electromagnetic directed-energy pulse trains" (EDEPTs)[26]. Subsequently, A wide variety of pulses have been established within the EDEPT family, including the modified power spectrum pulse[28], pulse with azimuthal dependence[29], linearly polarized pulse "focused pancakes"[30] and the toroidal symmetry "focused doughnut" (FD) pulse[28], etc. Among them, FD is attracting growing attention[31-35] owing to its skyrmionic field configurations[31,36], multiple vector singularities[37], and unusual space-time nonseparability[25,38]. In addition to above-mentioned spatiotemporal structured light pulses, the spatiotemporal optical vortices, such as spatiotemporal bessel beams, ultrafast X vortices, and transverse spatiotemporal vortices, also have been theoretically studied and experimentally constructed[39-41]. However, no classical entangled optical vortex pulses with space-time nonseparability has yet been reported.

In this paper, we theoretically propose a new kind of space-time nonseparable electromagnetic vortices (STNEV). The STNEV with abitrary OAM and multiple singularities displays spiral-like arrangements of the transient electromagnetic fields and the Poynting vector of the pulses indicates energy backflow phenomenon. In particular, we apply quantum measurements to quantitatively characterize the evolution of the pulse's structure upon propagation. Calculated results show that STNEV is space-time nonseparable with much greater entanglement characteristic parameters (fidelity, concurrence, and entanglement of formation) than the LG beam, spatiotemporal Bessel-Gaussian beam, spatiotemporal bessel vortices, and ultrafast X vortices.

# Results

## Fields of space-time nonseparable optical vortices

Following the EDEPTs derivation method, to obtain the electric and magnetic fields for STNEV, we start with an appropriate scalar generating function $f(\mathbf{r},t)$ that satisfies Helmholtz's wave equation[26] $\left(\nabla^2 - \frac{1}{c^2}\frac{\partial^2}{\partial t^2}\right)f(\mathbf{r},t) = 0$, where $c = 1/\sqrt{\mu_0\varepsilon_0}$ is the speed of light, and $\varepsilon_0$ and $\mu_0$ are the permittivity and permeability of the medium, respectively. Next, the exact solution of $f(\mathbf{r},t)$ can be given by the combination of modified power spectrum method and spiral phase factor[29], as $f(\mathbf{r},t) = \left(\frac{\rho}{q_1 + i\tau}\right)^{|l|} e^{il\theta} \frac{f_0}{\rho^2 + (q_1 + i\tau)(q_2 - i\sigma)}$, Where $\tau = z - ct$, $\sigma = z + ct$, $f_0$ is a normalizing constant and $l$ is a constant defining the topological number or the order. When compared to a Gaussian beam, the parameters $q_2$ and $q_1$ represent respectively the "Rayleigh range" (or depth of the focal region) and effective wavelength[28]. In particular, the value of the ratio $q_2/q_1$ indicates whether the pulse is strongly focused or collimated ($q_2/q_1 \gg 1$). The electromagnetic fields for the transverse electric (TE) solution can be derived by the potential $\Pi = \nabla \times \hat{z} f(\mathbf{r},t)$ as $E(\mathbf{r},t) = -\mu_0 \frac{\partial}{\partial t} \nabla \times \Pi$ and $H(\mathbf{r},t) = \nabla \times (\nabla \times \Pi)$. Finally, in a cylindrical coordinate system, the TE field components are given by the expressions:

$$E_\rho = \frac{f_0 l e^{il\theta}}{\rho}\sqrt{\frac{\mu_0}{\varepsilon_0}}\left(\frac{\rho}{q_1 + i\tau}\right)^{|l|}\left\{\frac{(q_2 + q_1 - 2ict)}{\left[\rho^2 + (q_1 + i\tau)(q_2 - i\sigma)\right]^2} + \frac{|l|}{\left[\rho^2 + (q_1 + i\tau)(q_2 - i\sigma)\right](q_1 + i\tau)}\right\} \quad (1)$$

$$E_\theta = if_0\sqrt{\frac{\mu_0}{\varepsilon_0}}e^{il\theta}\left(\frac{\rho}{q_1 + i\tau}\right)^{|l|}\left\{\frac{l^2}{\left[\rho^2 + (q_1 + i\tau)(q_2 - i\sigma)\right]\rho(q_1 + i\tau)} + \frac{|l|(q_2 + q_1 - 2ict)}{\rho\left[\rho^2 + (q_1 + i\tau)(q_2 - i\sigma)\right]^2} - \frac{4\rho(q_2 + q_1 - 2ict)}{\left[\rho^2 + (q_1 + i\tau)(q_2 - i\sigma)\right]^3} - \frac{2|l|\rho}{\left[\rho^2 + (q_1 + i\tau)(q_2 - i\sigma)\right]^2(q_1 + i\tau)}\right\} \quad (2)$$

$$H_\rho = f_0 e^{il\theta}\left(\frac{\rho}{q_1 + i\tau}\right)^{|l|}\left\{\frac{-il^2}{\left[\rho^2 + (q_1 + i\tau)(q_2 - i\sigma)\right]\rho(q_1 + i\tau)} + \frac{4\rho\left[(q_2 - q_1)i + 2z\right]}{\left[\rho^2 + (q_1 + i\tau)(q_2 - i\sigma)\right]^3} - \frac{\left[(q_2 - q_1)i + 2z\right]|l|}{\rho\left[\rho^2 + (q_1 + i\tau)(q_2 - i\sigma)\right]^2} + \frac{2i\rho|l|}{\left[\rho^2 + (q_1 + i\tau)(q_2 - i\sigma)\right]^2(q_1 + i\tau)}\right\} \quad (3)$$

$$H_\theta = \left(\frac{\rho}{q_1 + i\tau}\right)^{|l|} l f_0 e^{il\theta}\left\{-\frac{i\left[(q_2 - q_1)i + 2z\right]}{\left[\rho^2 + (q_1 + i\tau)(q_2 - i\sigma)\right]^2} + \frac{|l|}{\left[\rho^2 + (q_1 + i\tau)(q_2 - i\sigma)\right]\rho(q_1 + i\tau)}\right\} \quad (4)$$

$$H_z = 2f_0 e^{il\theta}\left(\frac{\rho}{q_1 + i\tau}\right)^{|l|}\left\{-\frac{2}{\left[\rho^2 + (q_1 + i\tau)(q_2 - i\sigma)\right]^2} + \frac{(q_2 + q_1 - 2ict)^2 + \left[(q_2 - q_1)i + 2z\right]^2}{\left[\rho^2 + (q_1 + i\tau)(q_2 - i\sigma)\right]^3} + \frac{|l|\left[(q_2 + q_1 - 2ict) + i\left[(q_2 - q_1)i + 2z\right]\right]}{\left[\rho^2 + (q_1 + i\tau)(q_2 - i\sigma)\right]^2(q_1 + i\tau)}\right\} \quad (5)$$

where $E_\rho$ and $E_\theta$ represent the radially and azimuthally directed component of electric field, and $H_\rho$, $H_\theta$ and $H_z$ are the radially, azimuthally and longitudinally directed components of magnetic field, respectively. Note that the TE mode field does not possess longitudinally directed components of electric field $E_z$. The transverse magnetic (TM) mode can be expressed by exchanging the electric and magnetic fields. For $l=0$, the electromagnetic fields in Equations (1)-(5) are reduced to the fundamental FD pulse, which has only three field components $E_\theta$, $H_\rho$ and $H_z$, because the generating function $f(\mathbf{r}, t)$ is assumed to be independent of $\theta$. Moreover, the real and imaginary parts of Equations (1)-(5), simultaneously satisfy Maxwell equations. See detailed derivation of Equations (1)-(5) in Supplementary Note 1.

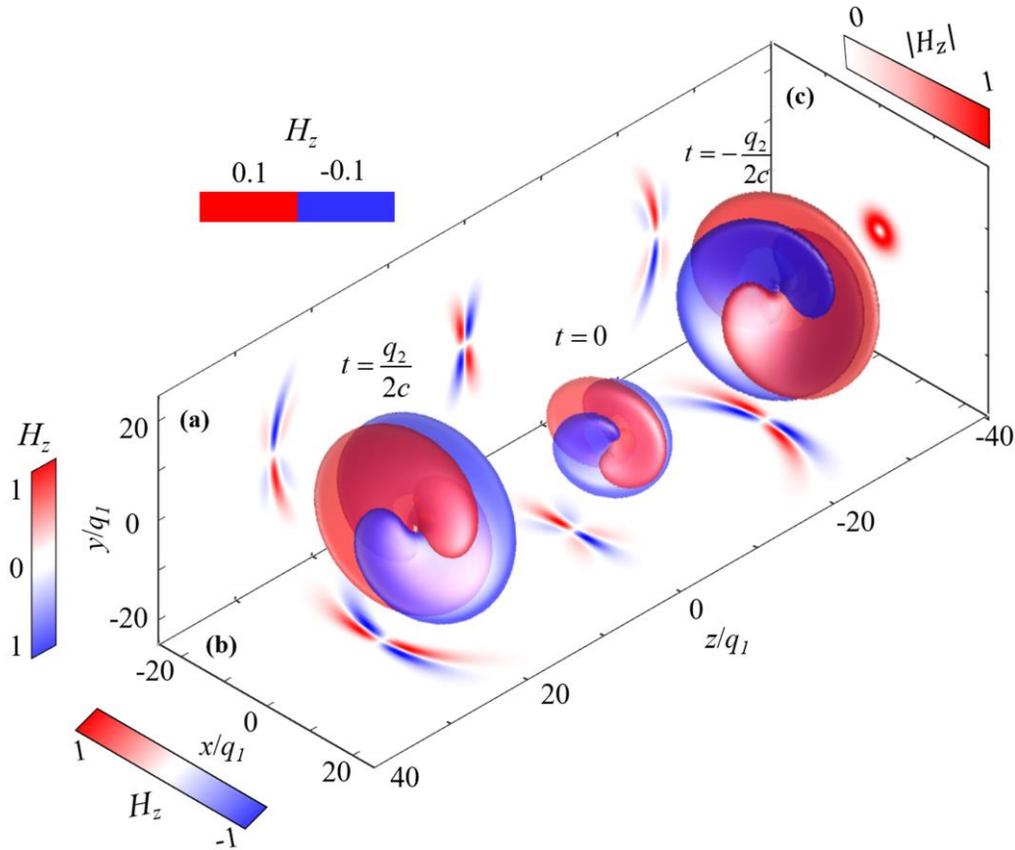

**Figure 1 Spatiotemporal structure of TE STNEV**: The spatial isosurfaces of the magnetic field Re($H_z$) at normalized amplitude levels of Re($H_z$)=±0.1 and the Rayleigh range of $q_2=50q_1$ and parameter $l=1$. **a** The $y$-$z$ cross-section of the instantaneous magnetic field Re($H_z$) at $x=0$ for $t=0$ and $\pm q_2/(2c)$. **b** The $x$-$z$ cross-section of the instantaneous magnetic field Re($H_z$) at $y=0$ for $t=0$ and $\pm q_2/(2c)$. **c** The $x$-$y$ cross-section of the magnetic field $H_z$ intensity at the $z=0$ plane.

While propagating in free space, the propagation dynamics are revealed by the isosurfaces of the real longitudinal magnetic field at various times. Figure 1 shows the evolution of the STNEV ($l$=1) upon propagation through the focal point. Combined with spiral-like and doughnut-like structure, the final topology structure of STNEV appears as two spiral donuts lobes twisted together, with positive and negative pulses distinguished in blue and red in Figure 1, respectively. Similar to Gaussian pulses, STNEV can be focused due to the Gouy phase shift[42]. The STNEV exhibit a complex spatiotemporal evolution like the fundamental FD, with the pulses being reshaped several times during propagation. See Supplementary Movies 1 and 2 for dynamic evolutions upon propagation of the STNEV, and Supplementary Note 4 for time-frequency distribution upon propagation of the STNEV.

To date, exact closed-form expressions for the STNEV have only been defined in the time domain. Although the time domain expressions allow to describe the STNEV, frequency domain expressions are crucial for understanding the propagation dynamics and examining the specture phase of STNEV. We have derived analytically expressions for the real part of $l$=0 and 1 TE STNEV. See detailed derivation of Equations in Supplementary Note 2. And based on the derived frequency domain expressions, we can find an intriguing feature of STNEV involving the presence of vortex phases, which can be observed in the intensity and phase distributions of STNEV and typical vortex Laguerre–Gaussian (LG) beams of different orders in Figure 2.

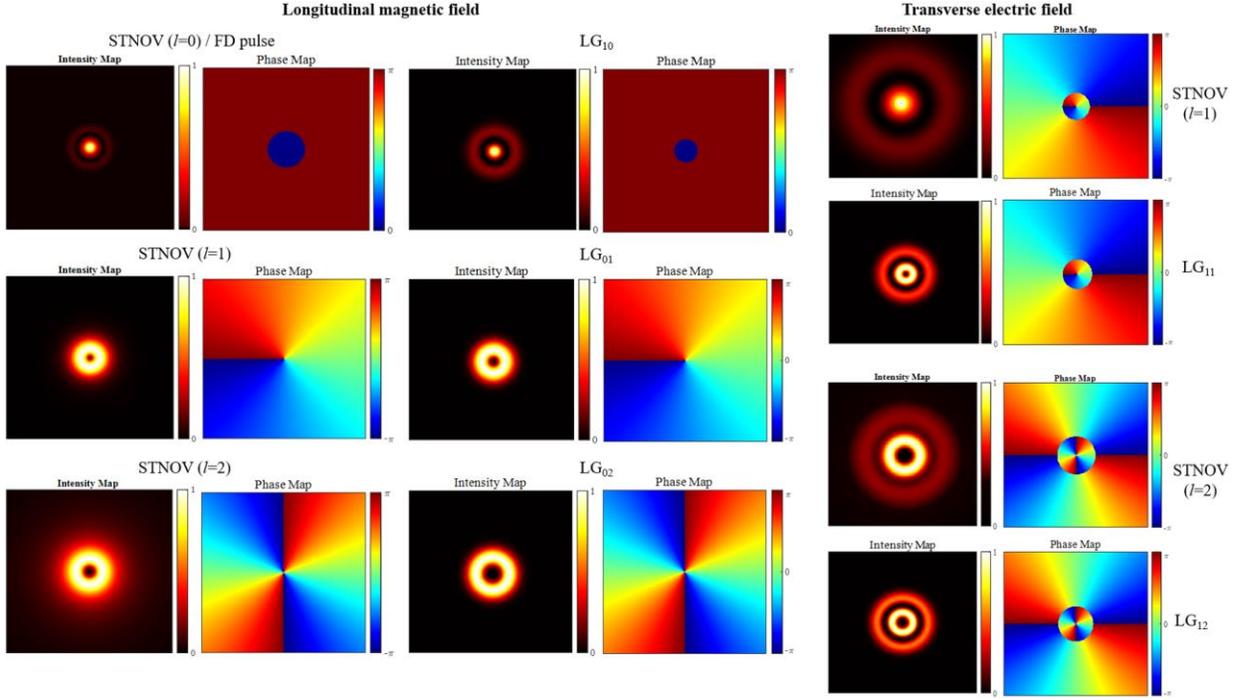

**Figure 2 Intensity and phase distribution of TE STNEV and LG beam with different orders**: with Rayleigh range $q_2=50q_1$ in the *x-y* plane at certain frequency.

From the figure, we can see that the intensity and phase distribution of the longitudinal magnetic field of STNEV with $l$=0, 1, and 2 correspond to the $LG_{10}$, $LG_{01}$, and $LG_{02}$ modes in typical vortex beams, respectively, while the intensity and phase distribution of the transverse electric field of STNEV with $l$=1 and 2 correspond to the $LG_{11}$ and $LG_{12}$ modes, respectively, which also proves that STNEV do have vortex phase and carry orbital angular momentum. We can also summarize the intensity and phase variations of STNEV as the topological charge *l* varies. The existence of a phase singularity in the center causes the light intensity to show a ring distribution with a central dark core. The larger the value of the topological charge *l*, the larger the diameter of the dark spot in the center of the intensity field, the gradually larger the diameter of the central bright ring, and the farther the location of the field intensity maximum is from the center. The phase changes by $2\pi l$ per cycle, and the phase distribution is also closely related to the number of light rings, because the phases of different bright rings are not synchronized, and sudden phase changes may occur.

## Space-time nonseparablity measurement

Since classical nonseparable pulse states are mathematically analog to quantum entangled states, based on this similarity, it is almost natural to exploit quantum methodology to quantitatively characterize the corresponding properties of classically structured pulse[44-47]. In the meantime, the structured pulse has developed to harness multiple complex degrees-of-freedom (DoFs) in itself, and new quantum-analog measure methods were modified to a high-dimensional state[48], we can apply an analogous method to a more complex structured pulse. For instance, the quantum state tomography measurement can be applied to vector vortex beams[49], to reconstruct the density matrix for characterizing the vector vortex beams and calculating the fidelity to reflect the quality to the ideal light state, akin to a measure of qubit. Over the past century, an extended toolbox has been developed to quantify the quality of a quantum entangled state and characterize the spatiotemporal propagation dynamics, including state tomography, density matrix, fidelity, concurrence, etc[50,51].

Typically, electromagnetic spatiotemporal pulses can be described by a bipartite state with the space and time DoFs, which can be expressed as a product of a spatial mode and a temporal (spectral) function, corresponding to a general high-dimensional bipartite state[10]. However, "electromagnetic directed-energy pulse trains" (EDEPTs)[26] as space-time nonseparable exact solutions of Maxwell's equations, whose family includes pulses such as FDs[28], are typical examples of STNS (space-time nonseparability) pulses, that cannot be expressed as products of spatial and temporal functions, corresponds to the maximally entangled nonseparable state[10]. Since vortex beams carrying helical phase, as classical analog quantum states associated with orbital angular momentum (OAM)[6], can also be represented as space-polarization nonseparable states, we believe that the new electromagnetic pulses proposed by introducing vortex phase factors in FD pulse still have space-time nonseparability, and we quantize STNS with measures of fidelity, coherence, and entanglement of formation in quantum measurements[52]. First, we can introduce two sets of states to describe STNS in pulses[25]: (1) *Spectral states* $|\lambda_i\rangle (i=1,2,\ldots,n)$ are states of light of defined wavelength $\lambda_i$ and with defined radial position ($r_{\lambda_i}$) of peak intensity; (2) *spatial states* $|\eta_i\rangle$ are states of light located at the position with a defined radial ratio of $\eta_i = r/r_{max}$, where $r_{max}$ is the radial position at which the total intensity of the light field reaches its maximum. Based on the prior theory[53], the introduction of spatial and spectral sets of states allows us to distinguish similar broadband waves. As an example, we consider six pulses with different STNS, a wideband LG beam, a fundamental FD, a STNEV ($l$=1), a spatiotemporal Bessel-Gaussian beam, a spatiotemporal bessel vortices and a ultrafast X vortices.

As illustrated by the corresponding spatial and spectral states, all pulses exhibit very different spatial-spectral structure and propagation dynamics. For the FD pulse and the STNEV, the spectral states are coincident with the corresponding spatial states upon propagation, as Figure 3b1 and c1 show. In contrast, the wideband LG beam, two spatiotemporal Bessel beams and the ultrafast X vortices's corresponding spectral and spatial states are naturally separated (see Figure 3a1 and d1-f1), and the spatial-spectral structure of these beams varies dramatically as they propagate. The difference among two type of beams can be emphasized further in the $\eta$–$z$ plane. Here, as expected, the spectral states of the FD pulse and the STNEV(Figure 3b2 and c2) are $z$ invariant and coincident with the corresponding spatial states. The crucial characteristic induced by the space-time nonseparable pulse is the isodiffraction nature, whereby the spatial distribution of different spectral components in the transverse plane does not suffer distortion upon propagation[33]. On the other hand, the profile of the remaining four beams (Figure 3a2 and d2-f2) suffer substantial distortion as illustrated by the trajectories of the spectral states. This is a direct result of the noncoincidence of spectral and spatial states.

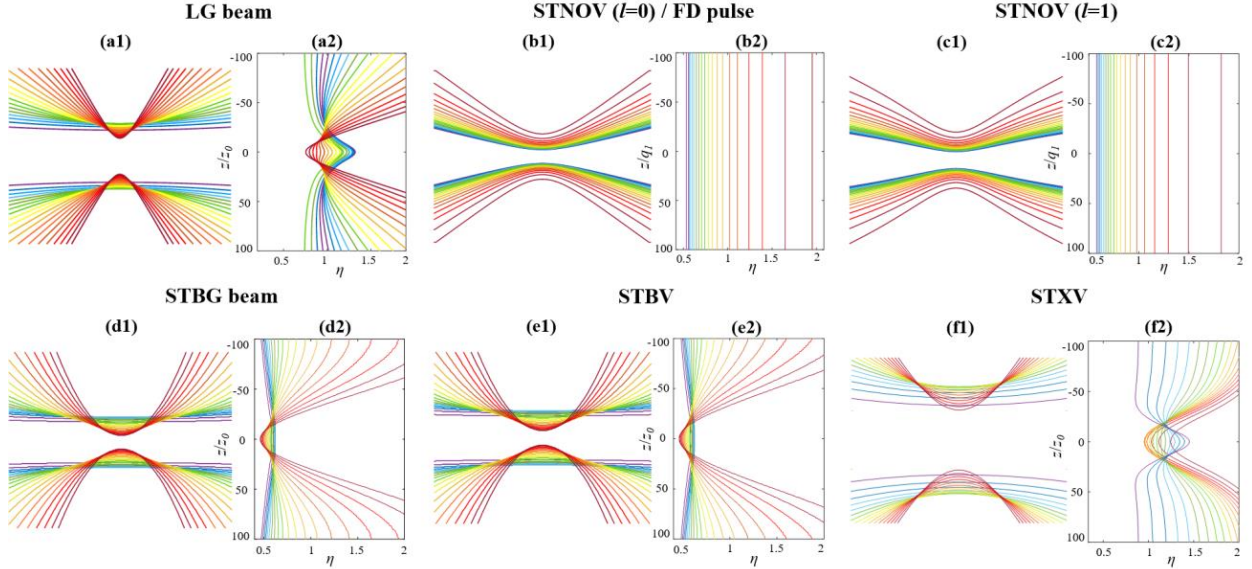

**Figure 3 The propagation profiles of spectral ($|\lambda_i\rangle$), spatial ($|r_i\rangle$) states and the $\eta$-$z$ map of different pulses**: the wideband LG beam (**a1-a2**), the FD pulse (**b1-b2**), the $l$=1 STNEV (**c1-c2**), the spatiotemporal Bessel-Gaussian beam

(**d1-d2**), the spatiotemporal bessel vortices (**e1-e2**) and the ultrafast X vortices (**f1-f2**). In this illustration, we have considered 20 different spectral and spatial states, $|\lambda_i\rangle$ and $|\eta_i\rangle$ ($i$ = 1, 2, ..., 20), where radial ratios are selected as $\eta_i = r_{\lambda_i}/r_{max}$.

To quantify STNS in pulse, the problem can be viewed as a measurement of an entanglementlike system, i.e., the nonseparability of two classical spatial fields. The classical fields of the spectral state $|\lambda_i\rangle$ and the spatial state $|\eta_i\rangle$ are represented as $\mathcal{E}_{\lambda_i}(r,z) = \sqrt{I(\lambda_i,r,z)}H(r-\delta_{i-1}^{(\lambda)})H(\delta_i^{(\lambda)}-r)$, $\mathcal{E}_{\eta_i}(r,z) = \sqrt{I_0(r,z)}H(r-\delta_{i-1}^{(\eta)})H(\delta_i^{(\eta)}-r)$, where $H(r)$ is the Heaviside step function, $H(r)=1$ if $r>0$ and is zero otherwise. Both sets of spectral and spatial states are orthogonal, $\langle\lambda_i|\lambda_j\rangle = \delta_{ij}$ and $\langle\eta_i|\eta_j\rangle = \delta_{ij}$, where $\delta_{ij}$ is the Kronecker delta. The inner product of two states is given by $\langle\eta_i|\lambda_j\rangle = \int \mathcal{E}_{\eta_i}\mathcal{E}_{\lambda_j}^* dr$. Here, the definitions of the classical fields $\mathcal{E}_{\lambda_i}, \mathcal{E}_{\eta_i}$ have been introduced with respect to the radial coordinate $r$. Based on the similar mathematical description and physical origin of the above two states, we then introduce the new concept of a *space-spectrum nonseparable state* that allows us to quantitatively describe pulses with prescribed STNS, for a general pulse, the space-spectrum state is expressed as $|\psi\rangle = \sum_{i}^{n}\sum_{j}^{n} c_{i,j}|\eta_i\rangle|\lambda_j\rangle$, where $c_{i,j}^2 = \langle\eta_i|\lambda_j\rangle$ and $\sum_{ij} c_{ij}^2 = 1$. Based on the definition of the spectral and spatial states adopted here[25], we can calculate the tomography matrix, and then we can reconstruct the corresponding density matrices of the space-spectrum state $\tilde{\varrho} = |\tilde{\psi}\rangle\langle\tilde{\psi}|$, (where $|\tilde{\psi}\rangle$ is the measured state). Importantly, knowledge of the density matrix allows us to apply quantum measures, e.g., fidelity and concurrence, to quantitatively characterize the properties of the pulse.

*Fidelity.* In quantum mechanics, the fidelity is a measure of similarity of two quantum states, defined as $F = \left(Tr\sqrt{\sqrt{\rho_1}\rho_2\sqrt{\rho_1}}\right)^2$, where $\rho_1$ and $\rho_2$ are the density matrices of the two states[50]. Here, we set the target state as the ideal FD pulse $|\psi\rangle = \sum_{i=1}^{n} c_i |r_i\rangle|\lambda_i\rangle$. The fidelity of a measured state can then be calculated as $F = \langle\tilde{\psi}|\tilde{\varrho}|\tilde{\psi}\rangle$, where $\tilde{\varrho} = |\tilde{\psi}\rangle\langle\tilde{\psi}|$ is the density matrix of the measured state. In our implementation, fidelity can quantitatively measure the degree of similarity to an ideal pulse taking values from 0 to 1, which indicates the degree of similarity to the ideal pulse.

*Concurrence.* In quantum mechanics, the concurrence is a continuous measure of nonseparability of two-dimensional entangled states, defined by $C = \sqrt{2\left(1-Tr\left(\rho_A^2\right)\right)}$, where $\rho_A$ is the reduced density matrix[57]. For an arbitrary $d$ dimensional state, the concurrence is usually normalized and takes values from 0 to 1( $v_d = \sqrt{2(1-1/d)}$ ), indicating the absence of entanglement (or pure separability) and strong nonseparability (maximum entanglement), respectively. In our study, we use $d$=20 corresponding to the 20 spectral and spatial states. The corresponding results for the ideal FD and STNEV show that both pulses exhibit strong STNS with near-maximum "entanglement".

*Entanglement of formation.* In quantum mechanics, the entanglement of formation (*EoF*) is also a commonly encountered measure of quantum entanglement[58]. *EoF* is calculated as $E = -Tr[\rho_A \log_2(\rho_A)]$ and is typically normalized as $E/\log_2(d)$ in the $d$-dimensional case. The high *EoF* for the FD and STNEV pulses reveals that they both exhibit a strong STNS.

We summarize the results of fidelity, concurrence and entanglement measures in Table 1, for normalized intensity measurements, the detailed theory and calculations are discussed in Supplementary Note 3. The similarity between classical and quantum worlds allows us to apply the quantum tools to describe many of the properties of classical nonseparable states of light[55,59]. with an increasing number of intriguing higher-dimensional structured light modes (space-time nonseparable pulse) enabling the simulation of an increasing number of higher-dimensional quantum states, which allows the transfer of more quantum tools to classical fields[25]. The quantumlike measures introduced here also have a clear physical meaning related to the propagation dynamics of spatiotemporal pulses. In particular, the density matrix contains the full information of the correlation between the spectral and spatial states. The fidelity quantifies the similarity between two pulses. On the other hand, concurrence and *EoF* quantify the overlap of spectral states with spatial states. Space-time nonseparable pulses provide unusual and largely unexplored degrees of freedom in structuring light that is yet to be exploited, moreover, the space-time nonseparability can be employed for the construction of propagation-invariant spatiotemporal pulses[54-56], as well as the above-discussed isodiffracting single-cycle pulses[25]. As such, there is growing interest in the generation and control of high-quality space-time nonseparable pulses.

**Table 1 Parameter comparison of various kinds of pulse.** N-Fid., N-Conc., and N. EoF, fidelity, concurrence, and EoF in intensity-normalized measurement, respectively.

| Pulse | W.LG | FD | STNEV | STBG | STBV | STXV |
|---|---|---|---|---|---|---|
| **N-Fid.** | 0.0016 | 1 | 1 | 0.0011 | 0.0025 | 0.0028 |
| **N-Conc.** | 0.7079 | 0.9975 | 0.9977 | 0.7065 | 0.6973 | 0.7050 |
| **N-EoF** | 0.2233 | 0.9807 | 0.9857 | 0.2227 | 0.2185 | 0.2220 |

## Electric field singularities

The instantaneous electric field for the TE pulse is defined as $\mathbf{E} = E_\rho \boldsymbol{\rho} + E_\theta \boldsymbol{\theta}$, which combines both the radial and azimuthal components. Figure 4 comparatively shows the instantaneous electric fields for the TE single-cycle STNEV ($l=1$ and $l=2$) with $q_2 = 20q_1$ at the focus ($t=0$). In all cases, there are always central singularities on $z$-axis ($\rho = 0$, see vertical solid black lines) owing to the radial and azimuthal polarization, and due to the extra $E_\rho$ component, the electric field is not zero at $z = 0$.

For $l = 1$ case, the electric field possesses four shell-like singular surfaces symmetrically distributed along $z$ and $\rho$ axis. In the transverse plane at the center of the pulse (Figure 4a1), it shows an overall counterclockwise rotation forming a large vortex and two small vortices around three singularities along the propagation axis. At a distance of $1q_1$ from the pulse center (Figure 4a3), the electric field still forms three vortices and corresponds to each of the three singularities, but at this time all change to a clockwise rotation direction. However, at the $z = 0.6q_1$ transverse plane of the singularity (Figure 4a2), the electric field no longer forms a whole rotating vortex, only two small clockwise vortices and three singularities are retained, and the remaining area is divided into four parts, each of which is pointed to outward the singularity.

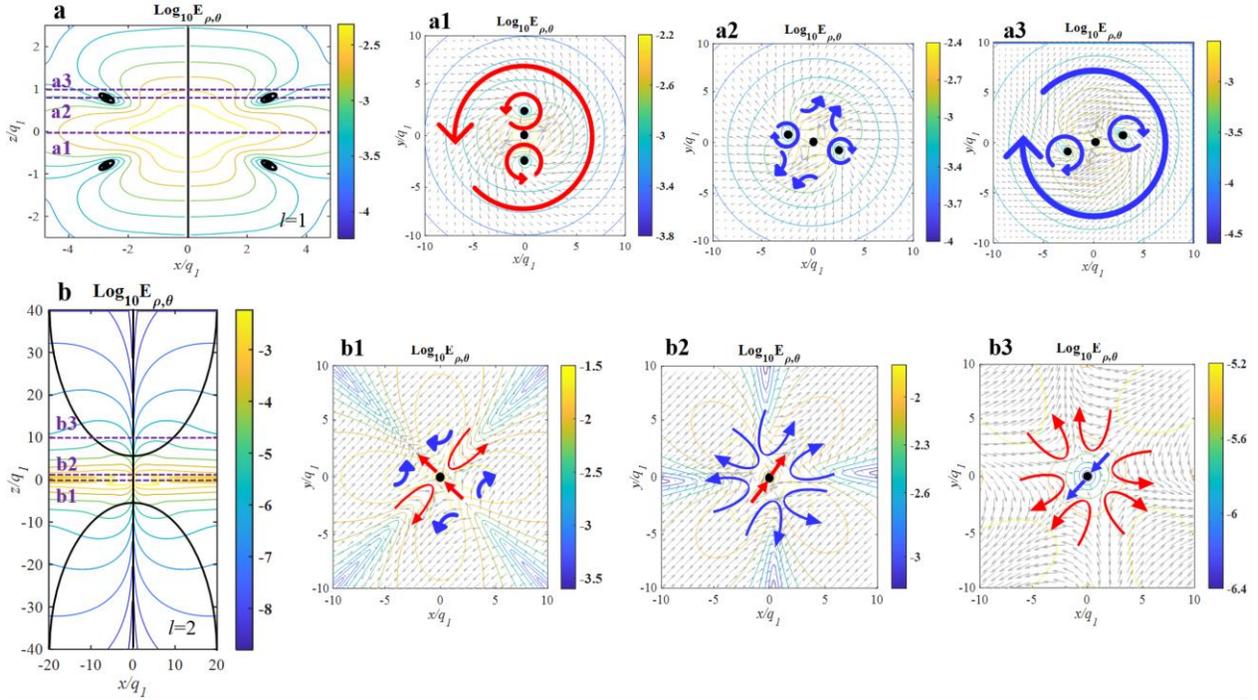

**Figure 4 Electric field topology of the single-cycle TE STNEV for different order *l* (a *l*=1 and b *l*=2) at focus.** The left panels present isoline plots of the electric fields in the *x*-*z* plane with logarithmic scale. The bold black lines represent the zero-value singular lines. The dashed purple lines represent the positions of propagation distance corresponding to the transverse plots on right side. Panels **a1**–**a3** and **b1**–**b3** show the single-cycle electric field distributions in the transverse *x*-*y* planes. The field magnitude is plotted as contours, while the field orientation is presented by arrow plots, where purple arrows also serve as a guide to the eye. Electric field zeros are marked by the black dots. The pulse is at focus (*t*=0) and propagates along the positive *z*-axis.

For $l = 2$ case, a more complex structure emerges with two shell-like singular surfaces symmetrically distributed along axis, as Figure 4b shows. The electric field configuration close to the singularity shells can be examined in detail at transverse planes at $z = 0q_1$, $0.6q_1$, $10q_1$ (Figure 4b1–b3). In the transverse plane of $z = 0$, the electric field space is divided into two large regions and four small regions, with the electric field rotating in opposite directions between the two large regions, while in the four small regions points from the central singularity to the periphery in an alternating clockwise and counterclockwise manner, respectively. At the plane away from the pulse $0.6q_1$, the electric field is divided into three different regions, each with two vortices rotating in opposite directions but pointing in the same direction, all pointed to the central singularity. While at the $z = 10q_1$ plane, the electric fields pointing in opposite directions comparaed to $z = 0.6q_1$ case. For the temporal evolution of the electric field singularities of the pulse see Supplementary Movies 3.

## Magnetic field singularities

The magnetic field of STNEV has radial, azimuthal and longitudinal three components, defined as $\mathbf{H} = H_\rho\boldsymbol{\rho} + H_\theta\boldsymbol{\theta} + H_z\hat{\mathbf{z}}$, which leads to presents an even more complex topological behavior than the electric field. Figure 5 comparatively shows the instantaneous magnetic fields for the STNEV of $l = 1$ and $l = 2$. For $l = 1$ case (Figure 5a), the magnetic field has ten different vector singularities on the $x$–$z$ plane, including two saddle points and two vortex-like rings on $z$-axis and six vortex rings (the surrounding vector distribution forming a vortex loop) away from the $z$-axis. The magnetic field distribution in the vicinity of the singularities is shown in more detail in Figure 5a1. Due to the axial symmetry of the pulse, the three singularities away from the $z$-axis correspond to three rings with vortices rotating clockwise ($z = 0$) or counter-clockwse ($z > 0$ and $z < 0$). Figure 5a2 presents the magnetic field distribution in the $z = 0$ plane, where the orientation of the magnetic field changes from $\rho > 0$ pointing to the positive $z$-axis to $\rho<0$ pointing to the negative $z$-axis, indicating that the magnetic field is antisymmetric with $\rho$=0 as the axis. For $l = 2$ case (Figure 5b), vector singularities are unveiled in the magnetic field with two saddle points on $z$-axis and six off-axis singularities. A zoom-in of the field structure around these singularities can be seen in Figure 5b1. In contrast to $l = 1$ case, the three singularities away from the $z$-axis correspond to three rings with vortices rotating counter-clockwise ($z = 0$) or clockwse ($z > 0$ and $z < 0$). The orientation of the magnetic field around the on-axis saddle points is alternating between "longitudinal-radial" and "radial-longitudinal". Moreover. Figure 5b2 presents the magnetic field distribution in the $z = 0$ plane, we can see that with $x$=0 and $y$=0 as the symmetry axes, respectively, the whole plane is divided into a four-part region, with the direction of the magnetic field in each part alternating in a parallel and anti-parallel $z$-axis. We can further deduce that the order $l$ determines the magnetic field distribution, the larger $l$ is, the more symmetry axes and divided regions the magnetic field distribution has. In all cases, we only consider the singularities existed at an area containing 99.9% of the energy of the pulse. While the singularity existed at the region far away from the pulse center with nearly zero energy can be neglected. For the temporal evolution of the magnetic field singularities of the pulse see Supplementary Movies 4.

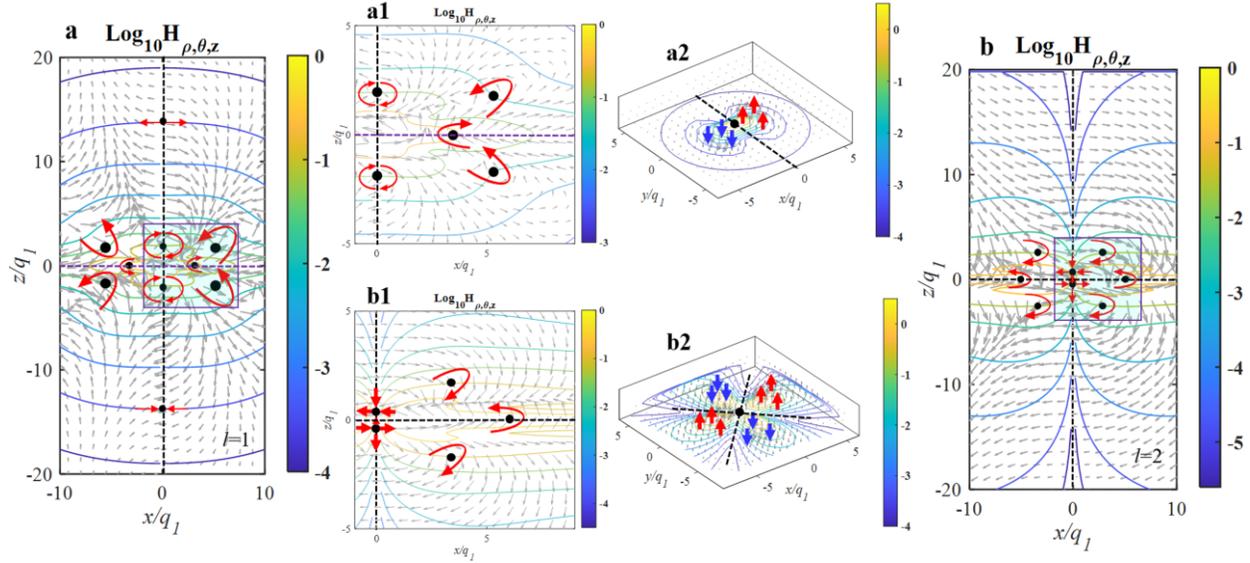

**Figure 5 Magnetic field topology of the single-cycle TE STNEV for different order *l* (a *l*=1 and b *l*=2) at focus**. in logarithmic scale, respectively. Magnetic field singularities are marked by black dots with red arrows correspondingly marking the saddle or vortex style of the vector singularities. Panels **a1** and **b1** present the zoom-in plots corresponding to regions of blue boxes in **a** and **b**, respectively. Panels **a2** and **b2** present transverse distributions of magnetic amplitude and normalized magnetic vectors in the *x-y* plane at *z*=0, the positions marked by the black dashed lines in **a** and **b**, respectively, where the magnetic fields with red arrows mark the singularities. The red arrows in all panels mark the direction of the magnetic field and serve as a guide for the eye. The pulse is at the focus (*t*=0) and propagates along the positive *z*-direction.

## Energy backflow and Poynting vector singularities

The topological features of electromagnetic fields in STNEV are linked to anomalous behavior of energy flow as represented by the Poynting vector $\vec{S} = \vec{E} \times \vec{H}$. An interesting effect is the presence of energy backflow: the Poynting vector at certain regions is oriented against the prorogation direction. Such energy backflow effects have been predicted and discussed in the context of singular superpositions of waves and plasmonic nanostructures[43]. The Poynting vector map reveals a complex energy backflow structure, as shown in Figure 6.

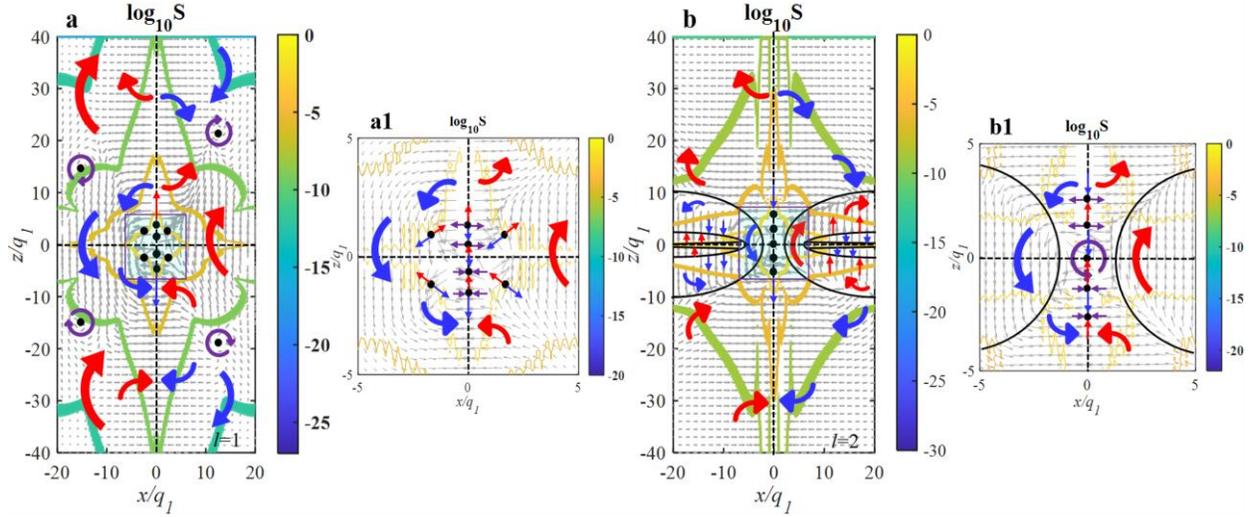

**Figure 6 Poynting vector of single-cycle TE STNEV at focus**. Contour and arrow plot of the logarithm of the Poynting vector of different order $l$ (**a** $l=1$ and **b** $l=2$) in the $x$-$z$ plane. Panels (**a1**) and (**b1**) present zoom-in of the areas highlighted by blue in (**a**) and (**b**), respectively. Solid black lines and dots mark the zeros of the Poynting vector. Red and bule bold arrows highlight the regions of energy forward flow and backflow, respectively.

In the case of $l=1$ (Figure 6a and a1), the energy flow vanishes along the propagation axis ($\rho=0$) and inherits the electric and magnetic singularities. For $l=2$ case (Figure 6b and b1), Poynting vector vanishes at along the $z$-axis and on the singular shells (marked by the black bold lines), presenting a more complex topological structure to that of $l=1$ pulse, this is a direct consequence of the more complex electric field topology of the pulse, which is also related to the temporal shape of the pulse. Importantly, the Poynting vector at regions of high intensity is oriented along the propagation direction (parallel to the positive $z$-axis; see red arrows), while at regions of relatively low intensity extended energy backflow can be observed (blue arrows), and, hence, the pulse as a whole still propagates forward. For the temporal evolution of the energy flow of the pulse see Supplementary Movies 5.

## Discussion

In this work, we show that STNEV exhibits complex and unique topological structure and energy backflow regions, which is related to the distribution of instantaneous electromagnetic and Poynting vector field of multiple saddle and vortex singularities. In addition to this, the space-time nonseparability is measured quantitatively by quantum measurement. Although our analysis focuses on a single ratio TE pulse, the variation of the scale does not affect its main features, and the findings can be easily extended to TM modes by simply exchanging the electric and magnetic fields.

The main challenges for the generation of space-time nonseparable electromagnetic vortices involve its complex toroidal topology, vortex phase and space-time nonseparable structure. We conjecture that the STNEV can be generated similarly to the generation of fundamental FD, by conversion of linearly polarized pulses in several-stage process. This process should include a conversion from linear to radial polarization, followed by a spatial-spectral correction of the pulse, and finally a phase adjustment to vortex phase.

In summary, STNEV combine both vortex phase with orbital angular momentum and space-time nonseparablity. We expect that an in-depth study of higher-order pulses will probably be the focus of future research, and it is highly probable that in the near future and with the exploitation of unexplored quantum-analog methods, other kinds of higher-dimensional nonseparable states of pulse with more DoFs will be discovered. Finally, our findings provide strong support for exploring STNEV's potential applications in areas including optical communications, energy transmission and microwave photonics.